\documentstyle[12pt]{article}
\catcode`\@=11
\catcode`\@=11

\def\section{\@startsection{section}{1}{\z@}{3.5ex plus 1ex minus
   .2ex}{2.3ex plus .2ex}{\large\bf}}

\def\ps@headings{\def\@oddfoot{}\def\@evenfoot{}
\def\@oddhead{\hbox{}\hfill
        \makebox[.5\textwidth]{\raggedright\ignorespaces --\thepage{}--
        \hfill }}
\def\@evenhead{\@oddhead}
\def\subsectionmark##1{\markboth{##1}{}}
}

\ps@headings

\catcode`\@=12
\catcode`\@=12

\relax

\def\figcap{\section*{Figure Captions\markboth
        {FIGURECAPTIONS}{FIGURECAPTIONS}}\list
        {Fig. \arabic{enumi}:\hfill}{\settowidth\labelwidth{Fig. 999:}
        \leftmargin\labelwidth
        \advance\leftmargin\labelsep\usecounter{enumi}}}
 \relax
\def\tablecap{\section*{Table Captions\markboth
        {TABLECAPTIONS}{TABLECAPTIONS}}\list
        {Table \arabic{enumi}:\hfill}{\settowidth\labelwidth{Table 999:}
        \leftmargin\labelwidth
        \advance\leftmargin\labelsep\usecounter{enumi}}}
 \relax
\def\reflist{\section*{References\markboth
        {REFLIST}{REFLIST}}\list
        {[\arabic{enumi}]\hfill}{\settowidth\labelwidth{[999]}
        \leftmargin\labelwidth
        \advance\leftmargin\labelsep\usecounter{enumi}}}
 \relax

\catcode`\@=11

\def\marginnote#1{}
\newcount\hour
\newcount\minute
\newtoks\amorpm
\hour=\time\divide\hour by60
\minute=\time{\multiply\hour by60 \global\advance\minute by-
\hour}
\edef\standardtime{{\ifnum\hour<12 \global\amorpm={am}%
    \else\global\amorpm={pm}\advance\hour by-12 \fi
    \ifnum\hour=0 \hour=12 \fi
    \number\hour:\ifnum\minute<100\fi\number\minute\the\amorpm}}
\edef\militarytime{\number\hour:\ifnum\minute<100\fi\number\minute}
\def\draftlabel#1{{\@bsphack\if@filesw {\let\thepage\relax
  \xdef\@gtempa{\write\@auxout{\string
    \newlabel{#1}{{\@currentlabel}{\thepage}}}}}\@gtempa
    \if@nobreak \ifvmode\nobreak\fi\fi\fi\@esphack}
     \gdef\@eqnlabel{#1}}
\def\@eqnlabel{}
\def\@vacuum{}
\def\draftmarginnote#1{\marginpar{\raggedright\scriptsize\tt#1}}
\def\draft{\oddsidemargin -.5truein
        \def\@oddfoot{\sl preliminary draft \hfil
        \rm\thepage\hfil\sl\today\quad\militarytime}
        \let\@evenfoot\@oddfoot \overfullrule 3pt
        \let\label=\draftlabel
        \let\marginnote=\draftmarginnote
\def\@eqnnum{(\theequation)\rlap{\kern\marginparsep\tt\@eqnlabel}%
\global\let\@eqnlabel\@vacuum}  }
\def\preprint{\twocolumn\sloppy\flushbottom\parindent 1em
        \leftmargini 2em\leftmarginv .5em\leftmarginvi .5em
        \oddsidemargin -.5in    \evensidemargin -.5in
        \columnsep 15mm \footheight 0pt
        \textwidth 250mmin      \topmargin  -.4in
        \headheight 12pt \topskip .4in
        \textheight 175mm
        \footskip 0pt

\def\@oddhead{\thepage\hfil\addtocounter{page}{1}\thepage}
        \let\@evenhead\@oddhead \def\@oddfoot{} \def\@evenfoot{}
}
\def\titlepage{\@restonecolfalse\if@twocolumn\@restonecoltrue\onecolumn
     \else \newpage \fi \thispagestyle{empty}\c@page\z@
        \def\thefootnote{\fnsymbol{footnote}} }
\def\endtitlepage{\if@restonecol\twocolumn \else  \fi
        \def\thefootnote{\arabic{footnote}}
        \setcounter{footnote}{0}}  
\catcode`@=12
\relax

\def\ps@headings{\def\@oddfoot{}\def\@evenfoot{}
\def\@oddhead{\hbox{}\hfill
        \makebox[.5\textwidth]{\raggedright\ignorespaces --\thepage{}--
        \hfill }}
\def\@evenhead{\@oddhead}
\def\subsectionmark##1{\markboth{##1}{}}
}

\ps@headings

\relax

\def\firstpage#1#2#3#4#5#6{
\begin{document}
\newcommand{\newc}{\newcommand}
\newc{\ra}{\rightarrow}
\newc{\lra}{\leftrightarrow}
\newc{\beq}{\begin{equation}}
\newc{\eeq}{\end{equation}}
\newc{\ba}{\begin{eqnarray}}
\newc{\ea}{\end{eqnarray}}
\newc{\bea}{\begin{eqnarray}}
\newc{\eea}{\end{eqnarray}}
\begin{titlepage}
\nopagebreak
\title{\begin{flushright}
        \vspace*{-0.8in}
{\normalsize math-ph/yymmddd}\\[-9mm]
\end{flushright}
\vfill
{#3}}
\author{\large #4 \\[1.0cm] #5}
\maketitle
\vskip -7mm
\nopagebreak
\begin{abstract}
{\noindent #6}
\end{abstract}
\vfill
\begin{flushleft}
{\normalsize IOA-TH/99-05}\\[-9mm] \vspace*{.6cm} {\normalsize May
1999}\\[-9mm]
\end{flushleft}
\thispagestyle{empty}
\end{titlepage}}

\def\simlt{\stackrel{<}{{}_\sim}}
\def\simgt{\stackrel{>}{{}_\sim}}
\date{}
\firstpage{3118}{IC/95/34} {\large\bf World volume Supermembrane
Instantons\\ in the light-cone frame.$^*$}
{E.G. Floratos$^{\,a,b}$ and G.K. Leontaris$^{\,c}$}
{\normalsize\sl
$^a$Institute of Nuclear Physics,  NRCS Demokritos,
{} Athens, Greece\\[-3mm]
\normalsize\sl
$^b$ Physics Department, University of Iraklion,
Crete, Greece\\[-3mm] \normalsize\sl
$^c$Theoretical Physics Division, Ioannina University,
GR-45110 Ioannina, Greece\\[-3mm] \normalsize\sl} {In this review
we present the octonionic duality for membranes. We start with a
discussion on the relation of the Yang Mills theories and the
supermembrane Hamiltonian in the light-cone gauge. We further
derive the self-duality equations for the membranes and discuss
the integrability of the system in 7 and 3 dimensions. Finally, we
present classical Euclidean time solutions of these equations and
examine the supersymmetries left intact by the self-duality
equations.
 } \vspace{2cm}
$^*$ Talk presented by E.G. Floratos in the workshop
``{\it Symmetries and Conservation Laws in Intermediate and High
Energy Physics }'', Ioannina, 1-5 October 1998

\newpage
\newpage
\section{Introduction}

One of the basic ingredients of $M$-threory\cite{w} is the eleven
dimensional ($11-d$) supermembrane for which some years
ago\cite{BST} a consistent action has been written in a general
background of the $11-d$ supergravity. The supermembrane has a
uniquely defined self-interaction which, in contrast to the
superstring, comes from a infinite dimensional gauge symmetry
apparent in the light-cone gauge as the area-preserving
diffeomorphisms ($SDiff\{M\}$) on the surface of the membrane.
There is no topological expansion over all possible three
manifolds analogous to the topological expansion of  the string
case,  because  of the absence of the dilaton field for the
supermembrane. The supermembrane, due to its unique
self-interaction, is possible to break into other supermembranes
so in a sense is already a second quantized theory
 but up to now there is no consistent perturbative
expansion around free harmonically oscillating supermembrane
configurations in topologically trivial space-time backgrounds. In
the light-cone gauge, and flat space-time, there are two classes
of membrane vacua, points and tensionless strings, so a low energy
effective field theory of supermembrane massless excitations would
be either eleven-dimensional supergravity or a field theory for
tensionless strings.

In a different approach, one may exploit the non-perturbative
structure of the supermembrane  vacuum in flat space--time
studying the corresponding Euclidean time equations of motion
which describe quantum tunneling between classical membrane
configurations\cite{BFS,fl,FZC}. The structure of quantum
mechanical vacua of supermembrane in flat backgrounds can be
studied through a new kind of self-duality\cite{fl} which is
apparent in the light-cone gauge as an analogue of
electric-magnetic non-abelian duality. Indeed, explicit
construction of the self-duality occurs in three\cite{fl}  and
seven\cite{FZC,flNP}  space dimensions due to the existence of
unique vector cross product defined by the quaternionic and
octonionic division algebras correspondingly. As in the case of
non-abelian duality, the corresponding (Euclidean space-time)
instantons  have a topological charge which has a space-time
interpretation for the membrane and  which satisfies a Bogomol'nyi
bound giving rise to supersymmetric invariance for the instantons.
In recent works, we have found explicit instanton solutions in
three and seven dimensions for the
$S^2$ and
$T^2$-membranes and we have shown that the corresponding surviving
supersymmetries are 8 and 1. The
$3-d$ instantons in principle can be constructed using Lax pair
techniques due to the integrability of this system. In seven
dimensions, there is an integrable sector but due to the
non-associativity of the octonionic algebra it seems that the
complete system is not integrable. In the following sections we
present a review on the derivation of the self-duality equations
and present some solutions. We will also discuss the
supersymmetries left unbroken by particular classes of solutions.
We note that analogous self-duality equations were written for Yang-Mills
(YM)-theories in \cite{bpst} and  described a  subclass of solutions
with interesting properties\cite{123}. Other generalizations of
such equations have also been intensively explored recently
\cite{tivan,popov,chap}.

\section{Matrix model and Membranes.}

It has been known since sometime that the supermembrane
Hamiltonian in the light-cone gauge is a very close relative of
YM-theories in the gauge $A_0=0$ and in one space
dimension less \cite{Hoppe}. To describe in some more detail this
relationship, we restrict our discussion to the bosonic part of
the Hamiltonian of the supermembrane in the light-cone gauge
 and to spherical topology for the membrane\cite{Hoppe,BST1,FIT}.
In reference\cite{FIT} it was pointed out that in the large $N$-limit,
$SU(N)$ YM theories have, at the classical level, a simple geometrical
 structure with the $SU(N)$ matrix potentials $A_{\mu}(X)$ replaced by
$c$-number functions of two additional coordinates $\theta, \phi$ of
an internal sphere $S^2$ at every space-time point,  while the
$SU(N)$ symmetry is replaced by the infinite dimensional algebra
of area preserving diffeomorphisms  of the sphere $S^2$ called
SDiff$(S^2)$. The $SU(N)$ fields ($N\times N$ matrices)
\begin{eqnarray}
& &A_{\mu}(X)=A^{\alpha}_{\mu}(X)t^{\alpha},\nonumber\\
& &t^{\alpha}\in SU(N),\nonumber\\
& &\alpha=1,2,\ldots,N^2-1,\quad \mu=0,1,\dots d-1 \label{1}
\end{eqnarray}
in the large $N$-limit become $c$-number functions of an internal
sphere $S^2$,
\begin{equation}
A_{\mu}(X,\theta,\phi)=\sum^{\infty}_{l=1}\sum^l_{m=-l}
A^{lm}_{\mu}(X)Y_{lm}(\theta,\phi),\label{2}
\end{equation}
where $Y_{lm}(\theta,\phi)$ are the spherical harmonics on $S^2$.
 The local gauge
transformations
\begin{equation}
\delta A_{\mu}=\partial_{\mu}\omega+[A_{\mu},\omega],\quad
\omega=\omega^{\alpha}t^{\alpha}, \label{3a}
\end{equation}
and
\begin{equation}
\delta F_{\mu\nu}=[ F_{\mu\nu},\omega],\label{3b}
\end{equation}
\begin{equation}
F_{\mu\nu}=\partial_{\mu}A_{\nu}-\partial_{\nu}A_{\mu}+
[A_{\mu},A_{\nu}] \label{lgt}
\end{equation}
are replaced by
\begin{equation}
\delta A_{\mu}(X,\theta,\phi)=\partial_{\mu}\omega
(X,\theta,\phi)+\{A_{\mu},\omega\},\label{4a}
\end{equation}
\begin{equation}
\delta F_{\mu\nu}(X,\theta,\phi)=\{F_{\mu\nu},\omega\},
\end{equation}
where
\begin{equation}
 F_{\mu\nu}(X,\theta,\phi)=\partial_{\mu}A_{\nu}-
\partial_{\nu}A_{\mu}+\{A_{\mu},A_{\mu}\},\label{5}
\end{equation}
and the Poisson bracket on $S^2$ is defined as follows:
\begin{equation}
\{f,g\}=\frac{\partial f}{\partial\phi}\frac{\partial g}
{\partial\cos\theta}-\frac{\partial g}{\partial\phi}
\frac{\partial f}{\partial\cos\theta} \label{6}
\end{equation}
So the commutators are replaced by Poisson brackets according to
\begin{equation}
\lim_{N\rightarrow\infty}N[A_{\mu},A_{\nu}]=\{A_{\mu},A_{\nu}\}
\label{7}
\end{equation}
Then the YM action in the large $N$-limit becomes \cite{FIT}
\begin{equation}
S_{\infty}=\frac{1}{16\pi g^2}\int_{S^2}d\Omega\int d^4XF_{\mu\nu}
(X,\theta,\phi)F^{\mu\nu}(X,\theta,\phi), \label{8}
\end{equation}
where
\begin{equation}
g=\lim_{N\rightarrow\infty}\frac{g_N}{N^{3/2}} \label{9}
\end{equation}
This large $N$-limit of $SU(N)$ YM-theories was found by making
use of the relation between the $SU(N=2 s+1)$ algebra in a
particular basis (up to spin $s$ $SU(2)-$tensor $N\times N$-
matrices) and
$SDiff(S^2)$ in the basis of the spherical harmonics
$Y_{lm}(\theta,\phi)$. In the present day language, this
$SDiff(S^2)$ YM-theory corresponds to the effective theory of
infinite number, $N\ra \infty$, $d-1$ dimensional Dirichlet
branes\cite{JP}. Here we note that the recently proposed matrix
theory which is claimed to be the long sought formulation of M
theory, is nothing but the SU(N) supersymmetric YM-mechanics which
was used as a consistent truncation of the
supermembrane\cite{Hoppe,Banks}.

The above considered large $N$-limit, is a very specific one which
depends on the appropriate basis of $SU(N)$ generators convenient
for the topology of the membrane and it has nothing to do, at
least in a direct way, with the planar approximation of
YM-theories. Also, it is different from the large $N$-limit used
in Matrix theory. In the case of the spherical membranes the
$SDiff(S^2)$ YM-theory describes the dynamics of an infinite
number of D0-branes forming a topological 2-sphere. In the light-cone
 gauge the transverse coordinates $X_i$, ($i,1...,9$) of the
$11-d$ bosonic part of the supermembrane satisfy the following
equations
 \beq \ddot{X}_i = \{X_k,\{X_k,X_i\}\}; \;\; i,k =
1,\ldots 9\label{eom}
 \eeq
 where summation over repeated indices is implied. The corresponding
 Gauss law which is the generator of the  $SDiff(S^2)$ group is given
 by the constraint
 \beq
\{X_i,\dot{X}_i\} = 0\label{GL}
 \eeq

From the point of view of Superstring theories the D0-branes are
point like solutions of low energy effective supergravity
description and the string excitations have to live in these
backgrounds. The superposition on the same point of $N$ D0-branes
still is a classical solution with its own space-time metric. The
supersymmetric $SU(N)$ matrix model appears to be the low energy
effective description of the excitations of this bound state and
the $SU(N)$ matrices which depend only on time, represent the
non-commutative position coordinates of these excitations. The
link with the old work of ref \cite{fl-fqm} about the
discretization of the membrane
 is the following\cite{afn}. The question was posed
in ref \cite{fl-fqm} about the geometrical meaning of the $SU(N)$
truncation of the area preserving  diffeomorphism symmetry of the
light-cone membrane Hamiltonian. The interpretation found was that
it is the algebra of canonical commutation Weyl-Heisenberg
relations for the momentum and position operators $P$ and $Q$
satisfying $QP=wPQ$ with $w=Exp(2 \pi\imath/N)$  which represent
the non-commuting coordinates of the discrete membrane surface
viewed as a discrete and finite $N\times N$ rectangular and
periodic phase space(toroidal membranes). The $N^2$ points of this
lattice represent due to non-commutativity $N$ degrees of freedom.
These are the $N$ D0-branes which make up the membrane in this
truncation. Finally the finite quantum mechanics developed
in\cite{fl-fqm,af} could be interpreted as single D0-excitations
with linear dynamics on the discrete membrane. According to this
interpretation the large $N$-limit of the $SU(N)$ supersymmetric
YM-quantum mechanics should provide the quantum membrane or an
infinite number of interacting strings
Up to now
only the infinite string picture has been discussed in the
literature with some convincing success \cite{Corfu98}. Perhaps
the sector of finite number of interacting strings provides
another cut for the membrane which could be seen as a field theory
of interacting strings, still in first quantized form.

\section{Self-dual membranes }

In this section we present a heuristic derivation of the
self-duality equations for the supermembrane and will show that it
can be formulated only in three and seven space dimensions. The
solutions of these equations, the supermembrane instantons are
characterized by a topological charge which is the degree of the
map from 3-parameter space of the membrane to the $3-d$ world
volume and which provides a lower bound for the Euclidean action,
in the same way as the instanton Bogomol'nyi bound for Yang-Mills
(YM) theories. We shall need only the bosonic part of the
light-cone Hamiltonian in flat space-time and the corresponding
constraint\cite{BST1}.

In order to explain the appearance of non-abelian electric-magnetic type
 of duality in the membrane theory, we recall that for YM-potentials
independent of space coordinates the self-duality equation in the gauge
$A_0=0$, is \cite{thooft,poly}
\ba
\dot{A}_i&=& \frac 12 \epsilon_{ijk}[A_j,A_k],\; i,j,k =
1,2,3 \label{sd3}
\ea
According to ref\cite{fcnd} the only non-trivial higher
dimensional YM self-duality equations exist in 8 space-time
dimensions which for the 7-space coordinate independent potentials
can be written (in the $A_0=0$ gauge) as
\ba
\dot{A}_i&=& \frac 12 \Psi_{ijk}[A_j,A_k],\; i,j,k = 1,\cdots 7
\label{sd7} \ea
 where $\Psi_{ijk}$ is the multiplication table of
the seven imaginary octonionic units\cite{guna}.

It is now tempting to take the large $N$-limit and replace the
commutators by Poisson brackets to obtain non-abelian (area
preserving diffeomorphisms) self-duality equations for membranes
in the light cone gauge for 3 and 7 space dimensions. In this
limit we replace the gauge potentials $A_i$ by the membrane
coordinates $X_i$. Then, the $3-d$ system is
 \ba \dot{X}_i&=&
\frac 12 \epsilon_{ijk}\{X_j,X_k\},\; i,j,k = 1,2,3 \label{sd3}
\ea while in seven space dimensions
 \ba
  \dot{X}_i&=& \frac 12
\Psi_{ijk}\{X_j,X_k\},\; i,j,k = 1,\cdots, 7 \label{sd7}
 \ea
It is easy to see that the self-duality membrane equations, imply
the second order Euclidean-time, equations of motion in the
light-cone gauge as well as the Gauss law. Indeed, the Gauss law
results automatically by making use of the $\Psi_{ijk}$ cyclic
symmetry \beq \{\dot{X}_i,X_i\}= 0 \eeq The Euclidean equations of
motion are obtained as follows
\ba
\ddot{X}_i& = & \frac
12\Psi_{ijk}\left(\{\dot{X}_j,X_k\}+ \{X_j,\dot{X}_k\}\right)\\
          & = & \{X_k,\{X_i,X_k\}\}
\ea
where use has been made of the identity
\beq
\Psi_{ijk}\Psi_{lmk}=\delta_{il}\delta_{jm}-\delta_{im}
\delta_{jl}+\phi_{ijlm}
\label{I1}
\eeq
and  of the cyclic property of the symbol
$\phi_{ijlm}$~\cite{guna}.


It has been noticed that the seven dimensional self-duality
equations can be appropriately presented using the octonionic or
Cauley algebra. The octonionic units $o_i$ satisfy the algebra
\beq
o_i o_j = -\delta_{ij} + \Psi_{ijk} o_k. \label{omult}
\eeq
where $o_i, i=1,\dots , 7$ are the 7 octonionic imaginary units
with the property
\beq \{o_i,o_j\} = - 2\delta_{ij},,
\eeq
 We choose the multiplication table
 \beq
\Psi_{ijk}=\left\{\begin{array}{ccccccc}1&2&4&3&6&5&7\\
                             2&4&3&6&5&7&1\\
                             3&6&5&7&1&2&4
 \end{array}\right.
\label{2.1}
\eeq
For later use, we note that the automorphism group of the
octonionic multiplication table is the exceptional group $G_2$. In
terms of these units an octonion  can be written as follows \beq
{X} = x_0 o_0 + \sum_{i=1}^7 x_io_i
\eeq with $o_0$ the identity element. The  conjugate octonion is
\beq
\bar{X} = x_0 o_0 -  \sum_{i=1}^7 x_io_i
\eeq

\section{ Integrability}

In this section we study the integrability of the self-dual
equations in 3 and 7 dimensions. Before doing this, we want to
factorize the time dependence of the membrane coordinates in order
to show the similarity of the system with the matrix Nahm
equations which are known to be integrable and  provide a general
method for the construction of the spherically symmetric BPS YM
monopoles\cite{rev}. Recent work of Fairlie and Ueno\cite{FU} has
shown that the 7-d generalization of the Euler-Top (a special case
of Nahm equations) provide an integrable system which has been
solved completely in terms of hyperelliptic functions\cite{Ueno}.
So, the time factorization of the self-dual membrane equations
provide integrable systems.

As in the case of the 3-d system\cite{fl} we may try to factorize
the time dependence in seven dimensions. We assume the following
factorization:
\beq X_i = Z_{ij}(t) f_j(\xi )
\eeq Then, from
eq.(\ref{sd7}),
we obtain
\bea \dot{Z}_{im}f_m& = & \frac
12\Psi_{ijk}Z_{jl}Z_{kn}\{f_l,f_n\}
\eea
We observe that if we
make the ansatz for the $7\times 7$ matrix
\beq \dot{Z}_{im}(t)
\Psi_{mln} = \Psi_{ijk} Z_{jl}(t)Z_{kn}(t) \eeq then the equation
\beq f_i = \frac{1}{2}\Psi_{ijk}\{f_j,f_k\} \label{feq}
\eeq
is automatically satisfied, while at the same time we have
succeeded in disentangling the time dependence from the
self-duality equation. Therefore, the problem is reduced to
finding solutions for $f_i(\xi)$ and $Z_{kl}$ equations
separately.

Another equivalent form of the previous equation for the matrices
$Z_{ij}$ is
 \bea
 \dot{Z}_{ij} &=& \frac
16\Psi_{ikl}\Psi_{jmn}Z_{km}Z_{ln}
 \eea
In the case of diagonal matrices $Z_{ij} = \delta_{ij} R_j(t)$, we
have
\bea
\dot{R}_i& =& \frac 16\Psi_{ikl}^2 R_k R_l
 \eea
  We now make  some observations
about  the symmetries of eqs(\ref{sd7}-\ref{feq}). If $X_i$ is a
solution of (\ref{sd7}), then for every matrix $R$ of the group
$G_2$, which is a subgroup of $SO(7)$,
\beq
Y_i = {\cal R}_{ij}
X_j\label{Rij}
\eeq
is automatically a solution of the same
equation, because the elements of $G_2$ preserve the structure
constants $\Psi_{ijk}$. In components,
 \beq
  \Psi_{ijk}{\cal R}_{kl}=
\Psi_{imn}{\cal R}_{mj}{\cal R}_{nl}\label{orth}
 \eeq
The above relation shows the  way to defining $G_2$ group elements
starting from two orthonormal seven-vectors. The equation is
obviously covariant under $SDiff(S2)$ transformations. One can
define combined $G_2$ and $SDiff(S2)$ transformations to get
 $SO(3)$ spherically symmetric solutions since $SO(3)$ can be realized
as a subalgebra of $SDiff(S2)$.

We note that in principle it is possible to look for non-linear symmetries
of the self-duality  equations, generalizing (\ref{Rij})
\beq
Y_i = f_i(X)
\eeq
where $f_i(X)$ must satisfy the equation
\beq
\Psi_{ijk}\frac{\partial f_k}{\partial X_l}= \Psi_{imn}\frac{\partial f_m}{\partial X_j}
\frac{\partial f_n}{\partial X_l}.
\eeq

 In the following we examine  the self-consistency of eq.(\ref{feq}).
 Multiplying by $\Psi_{ilm}$, we get
\beq
\Psi_{ilm} f_i = \{f_l,f_m\}+\frac 12\phi_{lmjk}\{f_j,f_k\}
\eeq
Then, since the Poisson brackets satisfy the Jacobi identity, the above equation
 is constrained to satisfy the identity
\beq
\frac 13\phi_{ijkl}f_l =\Psi_{ijm}\{f_m,f_k\}+ {\rm cyclic\; perm.\; of}\; (ijk).
\eeq
This system of  equations  is exactly the same as in (\ref{feq}).
Another check for the self-consistency of $f_i$ equations can be found
as follows.   Define the tensors
\beq
{X^{ij}}_{kl}(u) ={\Delta^{ij}}_{kl} +\frac u4{\phi^{ij}}_{kl},
\eeq
\beq
{X^{ij}}_{kl}(u) ={\Delta^{ij}}_{kl} +\frac u4{\phi^{ij}}_{kl},
\eeq
where ${\Delta^{ij}}_{kl}=\frac 12(\delta_k^i\delta_l^j-\delta_l^i\delta_k^j)$
and the symbol ${\phi^{ij}}_{kl}\equiv \phi_{ijkl}$.
Then,  eqs(\ref{feq})  can be written as follows
\beq
\Psi_{ijk}f_k = {X^{ij}}_{lm}(2)\{f_l,f_m\}.
\eeq
Using now the algebra of the ${X^{ij}}_{kl}(u)$ tensors discussed
in detail in the Appendix of ref\cite{flNP}, we can prove that
both the identities
\beq
\Psi_{ijk}{X^{jk}}_{lm}(-1) =0
\eeq
and
\beq
 {X^{ij}}_{mn}(-1){X^{mn}}_{kl}(2) = 0
\eeq
hold, and this terminates the second consistency check.

The $3-d$ self-duality system has a Lax pair and infinite number of
 conservation laws\cite{fl}.
In order to see this, first we rewrite eqs.$(\ref{sd3})$ in the
form
\begin{equation}
\dot{X}_+=i\{X_3,X_+\}, \;\;
\dot{X}_-=i\{X_3,X_-\}, \;\;
\dot{X}_3=\frac{1}{2}i\{X_+,X_-\}, \label{17c}
\end{equation}
where
\begin{equation}
X_{\pm}=X_1\pm iX_2 \label{18}
\end{equation}
The Lax pair  equations can be written as
\beq
\dot{\psi}=L_{X_3+\lambda X_-}\psi, \quad \dot{\psi}=L_{\frac{1}{\lambda}
X_{+}-X_3}\psi,\label{21a,b}
\eeq
where the differential operators $L_f$ are defined as
\beq
L_f\equiv i\big(\frac{\partial f}{\partial\phi}
\frac{\partial}{\partial\cos\theta}-\frac{\partial f}
{\partial\cos\theta}\frac{\partial}{\partial\phi}\big).\label{22}
\eeq
The compatibility condition of $(\ref{21a,b})$ is
\beq
[\partial_t-L_{X_3+\lambda X_-},\partial_t-L_{\frac{1}{\lambda}
X_{+}-X_3}]=0,\label{23}
\eeq
from which, comparing the two sides for the coefficients of the powers
$\frac{1}{\lambda},\lambda^0,\lambda^1$ of the spectral parameter
$\lambda$, we find $(\ref{17c})$. From the linear system  $(\ref{21a,b})$
using the inverse scattering method, one could in principle
construct all solutions of the self-duality equations.

The infinite number of conservation laws, are derived as follows
\cite{prep}:
{}From the Cartesian formulation
 \bea \frac{d X_i}{d t}&=& \frac
12 \epsilon_{ijk}\{X_j,X_k\} \eea contracting with a complex
3-vector $u_i$, such that \bea u_i &=& \epsilon_{ijk}u_jv_k \eea
where $u_iu_i = 0$, and $v$ another complex vector with
$v_iv_i = -1$ and $u_iv_i=0$,
we find, \bea \frac{d (u\cdot X)}{d t}&=& \{u\cdot X, v\cdot X\}
\label{3dsd} \eea The latter is a Lax pair type equation  which
implies \bea \frac{d }{d t}\int d^2 \xi (u\cdot X)^n = 0 \eea

Application of the same method after determining two complex 7-vectors
$u_i, v_i$ such that $u_iu_i = 0$,  $v_iv_i = -1$ and $u_iv_i=0$ leads
to the equation \bea \frac{d (u \cdot X)}{d t} &=& \{ u\cdot X,
v\cdot X\} + \frac 12\phi_{jklm} u_j v_k
                \{ X_l,X_m\}
\eea
The curvature tensor $\phi_{jklm}$ is defined as the dual of
$\Psi_{ijk}$ in seven dimensions. When the previous equation is
restricted to three dimensions we recover (\ref{3dsd}).

\section{Multistring solutions}

We have seen in the previous section that the self-duality
equations of the membrane are first order differential equations
which imply the second order --Euclidean time-- equations of
motion. These equations are non--linear and a general solution is
rather a hard task. It is possible however to work out
systematically membrane solutions which exhibit particular
symmetries\cite{fl,flpt,Ueno,FU,RW,flsu,CP}. Such cases are those with
spherical or axial symmetry, string like configurations etc. We
will present here few of the classical solutions of
refs\cite{fl,flpt}.  We start in this section with the
string--like solution of the self-duality equation (\ref{sd7}) in
7 dimensions. We assume the form
\beq X_i(\sigma_{1,2},t) = A_i
\sigma_1 +B_i \sigma_2 +P_i t+ f_i(\sigma_1,\sigma_2,t)
 \eeq
with $i=1,...,7$, and $f$ being a periodic function of
$\sigma_{1,2}$ and $A,B$ integer vectors.  Then we obtain
 \bea
 P_i & =&
\Psi_{ijk} A_j B_k\label{Pi}\\
\dot{f}_i & = & \Psi_{ijk}
\left(A_j \frac{\partial}{\partial\sigma_2}-B_j
 \frac{\partial}{\partial\sigma_1}\right)
  f_k +\frac{1}{2}\Psi_{ijk}\{f_j,f_k\}
\label{Pi1} \eea
Since $f$ is a periodic function with respect to
$\sigma_{1,2}$, we can expand it in terms of an infinite number of
strings, depending on the coordinate $\sigma_1$:
\beq
f_i(\sigma_1,\sigma_2,t) = \sum_nX_i^n(\sigma_1,t) e^{in\sigma_2}.
\eeq
Then, from the self-duality equations (\ref{Pi},\ref{Pi1})
 we find that the winding number of the
membrane is related to the center-of-mass momentum, which is
transverse to the compactification directions $A$ and $B$. Also,
the infinite number of strings are coupled through the following
equations
\beq
\dot{X}_i^n(\sigma_1,t) = \Psi_{ijk} \left(\imath n
A_j - B_j \frac{\partial}{\partial\sigma_1}\right) X_k^n +\frac{
\imath}{ 2}\Psi_{ijk} \sum_{n_1+n_2=n}\left(n_2 \frac{\partial
X_j^{n_1}}{\partial\sigma_1}X_k^{n_2}-n_1X_j^{n_1}
 \frac{\partial X_k^{n_2}}{\partial\sigma_1}\right)
\eeq The string-like solution corresponds to the particular case
${\partial f_i}/{\partial \sigma_2}=0$, where we obtain
 \beq
 X_i^0
= X_i(\sigma_1,t)\ra \dot{X}_i = \Psi_{ijk}B_k
 \frac{\partial
X_j}{\partial\sigma_1}.
 \eeq
 This equation is formally solved in
vector form by
\beq
 X (\sigma_1, t) = e^{t M \frac{\partial
}{\partial\sigma_1}} X (\sigma_1, 0) \label{form}
 \eeq
 where we defined the $7 \times 7$ matrix $M_{ij} = \Psi_{ijk}B_k$.

 Relation (\ref{Pi}) is now simply written $P_i=M_{ij}A_j$.
 On the other hand, explicit solutions of (\ref{form}) are found by
 expanding $X_i$ in terms of the eigenvectors of $M$. In fact, since
 $M$ is real and antisymmetric, the real 7-dimensional vector space
 decomposes into three orthogonal two-dimensional subspaces, each
 corresponding to a pair of imaginary eigenvalues $\pm \imath\lambda$,
 and a one-dimensional subspace, in the direction of $B_i$, corresponding
to the zero eigenvalue. Since, in addition, $(M^2)_{ij} = - B^2
\delta_{ij} + B_i B_j$ (as can be checked), we see that the
imaginary eigenvalue pairs are all $\pm \imath |B|$. Therefore the
problem decomposes into three 3-dimensional problems (one for each
subspace) of the kind we solved before. The general solution is
then \bea &X_1^{(n)} +\imath  X_2^{(n)} = F_n ( \sigma_1 -\imath B
t ) ~,~~~n=1,2,3\nonumber\\ &X^{(0)} = |B| t \nonumber\eea 
where $(X_1^{(n)} ,
X_2^{(n)} )$ are the projections of the membrane coordinates on
the $n$-th two-dimensional eigenspace and $X^{(0)} = X_i B_i /
|B|$ is the projection on $B_i$. As an example, if we choose $B_i$
in the third direction, $B_i = B \delta_{i3}$, we have
 \bea X_1 +
i X_2 = F_1 ( \sigma_1 -\imath B t ),
 &X_5 + i X_4 = F_2 (\sigma_1 -\imath B t )\nonumber\\
 X_6 + i X_7 = F_3 ( \sigma_1 -\imath B t),
&{\rm and} \;\;\;\;\;X_3 = B t.\nonumber
 \eea

In 3 dimensions  there is a variety of solutions. Restricting to
the axi-symmetric class of instanton like solutions, the
self-duality equations result to the continuous Toda equation
Indeed, the Ansatz
 \ba
  A_1  = R(\sigma_1,t) \cos \sigma_2 ,& A_2
= R(\sigma_1,t) \sin \sigma_2 ,& A_3 = z(\sigma_1,t)
\ea
leads to the axially symmetric continuous Toda  equation \beq
\frac{d^2\Psi}{d t^2}+ \frac{d^2{e^{\Psi}}}{d \sigma_1^2}=0,
\label{,.} \eeq where
$R^2 = e^{\Psi}$. Solutions of this equation have been discussed
in the literature in connection with the self-dual 4d Einstein
metrics with rotational and axial Killing vectors~\cite{RW}. Here
though, we note that $\sigma_1$ runs in a compact interval ($0,
2\pi$) for torus and ($-1,1$) for the sphere.

A specific example of a solution with separation of variables of
the Toda equation, in the case of spherical topology ($\sigma_1
=\cos\theta$, $\sigma_2=\phi$) is
 \ba R(\theta,t)&=&
\kappa\frac{\sin\theta}{\sinh\left[\kappa (t_0-t)\right]}\nonumber\\
z(\theta,t)&=&  \kappa \coth\left[\kappa (t_0-t)\right]\cos\theta,
\nonumber\ea
where $\kappa$ is a constant. A second solution can be obtained if
we change the hyperbolic functions with ordinary trigonometric
ones. A variation of the method of ref.~\cite{RW} where by
inversion of the non-linear system (\ref{sd3}) is finally
presented in ref\cite{flpt}.

\section{Supersymmetries}

In our analysis up to now we dealt only with the bosonic part of
the Hamiltonian. In this section  we will explore the number of
supersymmetries preserved by (3+1)- and (7+1)-dimensional
solutions. We will see that 3-d solutions preserve as many as
eight out of the sixteen supersymmetries while the 7-d
self-duality equations preserve only one supersymmetry\cite{flsu}.

There is a connection of the octonionic and Clifford algebra in
$d=8$ dimensions which may be used to study the octonionic
self-duality equations under supersymmetry transformations. The
relation with the octonions has been noticed in studying
compactifications of 11-d supergravity on $S^7$ as well as in the
case of $N=8$ gauged supergravities\cite{sg,ivan}. The embedding
of YM-instantons in 10-d has been constructed in\cite{hsip}.

The supersymmetry transformation is defined~\cite{BST}
\bea
\delta\theta &=&\frac 12\left(
\Gamma_+(\Gamma_I\dot{X}^I+\Gamma_-)+
                  \frac 12\Gamma_+\Gamma^{IJ}\{X_I,X_J\}\right)
\left(\begin{array}{c}
\imath\epsilon_A\\
\epsilon_B\end{array} \right)\nonumber
\eea
In terms of the $16\times 16$ $\gamma$-matrices, the above is written
\bea
\delta\theta &=& \left(\begin{array}{cc}
0 & 0\\
\imath\sqrt{2}\left(\gamma^I\dot{X}_I+\frac 12\gamma^{IJ}
\{X_I,X_J\}\right)&-2\cdot 1_{16}
\end{array} \right)\left(\begin{array}{c}
\imath\epsilon_A\\
\epsilon_B\end{array} \right)\label{delthe32}
\eea
which implies that
\bea
\sqrt{2}\left(\gamma^I\dot{X}_I+
\frac 12\gamma^{IJ}\{X_I,X_J\}\right)\epsilon_A
+ 2\cdot 1_{16} \epsilon_B=0\label{delthe16}
\eea
where $ \epsilon_A,  \epsilon_B$ are 16-dimensional spinors. {}From
the form of eq.(\ref{delthe16}), we  observe that if self-duality
equations are going to play a role in the preservation of a number of
supersymmetries, we should necessarily impose the condition $\epsilon_B=0$.
Thus, at least half of the supersymmetries are broken.  Now, the last term
in (\ref{delthe16}) is zero and eq.(\ref{delthe16}) simply becomes
\bea
\left(\gamma^I\dot{X}_I+
\frac 12\gamma^{IJ}\{X_I,X_J\}\right)\epsilon_A = 0\label{16SD}
\label{delthe16A}
\eea
Under the assumption that $\dot{X}_{8,9}=0$, it can be  shown that the above
reduces to a simpler --$8\times 8$-- matrix equation. In order to find a
convenient explicit form, we first express the $16\times 16$ matrices in
terms of the octonionic structure constants $\Psi_{ijk}$ as follows:
let the index $n$ run from 1 to 7; then we define
\bea
\gamma_{8} =
\left(\begin{array}{cc}
0 & 1_8 \\
-1_8&0
\end{array} \right),&
\gamma_{n} =
\left(\begin{array}{cc}
0 & \beta_n \\
-\beta_n&0
\end{array} \right) \label{gn}
\eea
where $1_8$ is the $8\times 8$-identity matrix and $\beta_n$
are seven  $8\times 8$  $\gamma$-matrices with
elements~\cite{ivan} \bea (\beta_n)^i_j = \Psi_{imj},&
(\beta_n)^i_8=\delta^i_j,
&(\beta_n)^8_j=-\delta^i_j
\label{belem} \eea
while it can be easily checked that
$\beta_1\cdots\beta_7=-1_8$ and \bea \gamma_9
&=& \left(\begin{array}{cc}
 1_{8}&0 \\
0&-1_8
\end{array} \right)
\eea
The commutation relations of $\beta_m$ give:
\bea
\left([\beta_m,\beta_n]\right)^8_j&=& +2 \Psi_{nmj}
\\
\left([\beta_m,\beta_n]\right)^j_8&=&-2 \Psi_{nmj}
\\
\left([\beta_m,\beta_n]\right)^i_j&=&-2{{\cal X}^{mn}}_{ij}(-4)
\eea
where the tensors ${{\cal X}^{mn}}_{ij}(u)$ are defined as follows~\cite{flNP}
\beq
{{\cal X}^{ij}}_{kl}(u) = {\Delta^{ij}}_{kl}+\frac u4{\phi^{ij}}_{kl}
\eeq
where ${\Delta^{ij}}_{kl}=\frac 12(\delta_k^i\delta_l^j-\delta_l^i\delta_k^j)$.
Next, we impose the following condition on the components of the 16-spinor
$\epsilon_A$
\bea
\epsilon_A&=& \left(\begin{array}{c}1\\
                                    -\imath
              \end{array} \right)
 \otimes
             \varepsilon\label{8spin}
\eea
where $\otimes$ stands for the direct product and
$\varepsilon$ is an eight-component spinor whose components are
left unspecified. Clearly, condition (\ref{8spin}) reduces further
the sixteen supersymmetry charges to eight. Separating the eight
components of
$\varepsilon= (\varepsilon_7, \varepsilon_1)$ where
$\varepsilon_{7(1)}$ is a seven-(one-) dimensional vector, we find
that eq.(\ref{16SD}) reduces to the matrix equation \bea
 {\cal O}\;\varepsilon\equiv\left(\begin{array}{cc}
\Psi_{imj}\dot{X}_m+\frac{\imath}2{{\cal X}^{mn}}_{ij}(-4)\{X_m,X_n\}
 & \dot{X}_i+\frac{\imath}2\Psi_{imn}\{X_m,X_n\}\\
-\left(\dot{X}_i+\frac{\imath}2\Psi_{imn}\{X_m,X_n\}\right)&0
\end{array} \right)\left(\begin{array}{c}
\varepsilon_7\\
\varepsilon_1\end{array} \right) = 0\label{delthe8}
\eea
The rather interesting fact here is that the matrix elements
${\cal O}_{8j}$ and ${\cal O}_{j8}$, $(j=1,..., 7)$ multiplying
the $\varepsilon_1$-component are  the self-duality equations
(\ref{sde}) in eight dimensions when the Euclidean time-parameter
$t$ is replaced with $\imath t$ (Minkowski).
Thus, $\varepsilon_1$-component remains unspecified and there is
always one supersymmetry  unbroken for any eight-dimensional
solution of the self-duality equations.

Let us now turn our discussion to the upper $7\times 7$ part
of the matrix equation (\ref{delthe8}). In general, the quantity
specifying these elements, namely
\beq
\Psi_{imj}\dot{X}_m+\frac{\imath}2{{\cal X}^{mn}}_{ij}(-4)\{X_m,X_n\}
\label{7x7}
\eeq
is {\it not} automatically zero. However, there is a particular case
--which turns out to be the most interesting one-- where the above
quantity is the self-duality equation itself. In fact, if we consider
only three-dimensional solutions of the equations, the `curvature'
factor ${\phi^{ij}}_{kl}$ is automatically zero while the tensor
${{\cal X}^{ij}}_{kl}$ simply becomes
\bea
{{\cal X}^{ij}}_{kl}=
{\Delta^{ij}}_{kl}=\frac 12(\delta_k^i\delta_l^j-\delta_l^i\delta_k^j)
&
{\rm for}\;\; {\phi^{ij}}_{kl}=0.
\eea
In this case, it can be easily seen that (\ref{7x7}) reduces to the
self-duality equations in three-dimensions. In this latter case, all
eight supersymmetries survive.

We conclude this short review with a few general remarks.
The  octonionic algebra  gives a useful formulation
of the self-duality equations, 
which includes in a natural way the three-dimensional
 system and the corresponding  generalized Nahm's equations for $SDiff S_2$.
By introducing in the place of the $SU(2)$ 
algebra of functions on the sphere, a
 quadratic
algebra of seven functions with $G_2$ symmetry, 
we succeeded to factorize the time
in a simple way, which 
may facilitate the study of solutions of the self-duality
 equations.
Although the general system of self-duality 
equations in seven dimensions does n
ot seem
to have a Lax pair, at least in a direct way, 
due to the non-associativity of the
octonionic algebra, it may happen that there 
is a generalization of the zero-curvature
condition under which this system is integrable.
We also explored the
supersymmetric self-duality configurations in three and seven
dimensions where we showed that in the 
three-dimensional integrable self-dual sector where
eight supersymmetries survive.

\end{document}